\documentclass[10pt,conference,english]{IEEEtran}
\usepackage[centertags]{amsmath}
\usepackage{amsfonts}
\usepackage{amssymb}
\usepackage{epsfig}
\usepackage{amsmath,  amssymb}
\usepackage{graphicx}
\usepackage[centertags]{amsmath}
\usepackage{clrscode}
\usepackage{cite}
\usepackage{fullpage}
\usepackage{amsmath,amssymb,mathrsfs,pseudocode}
\usepackage{color}
\usepackage[normalem]{ulem}
\usepackage{psfrag}
\usepackage{url}
\usepackage{babel}
\usepackage{authblk}

\usepackage[linesnumbered,ruled]{algorithm2e}
\usepackage[margin=18.35mm]{geometry}
\usepackage{xspace}
\usepackage{multicol}
\usepackage{booktabs} 
\usepackage{multirow}
\usepackage[x11names,dvipsnames,table]{xcolor}

\usepackage{subfigure}

\usepackage{stackrel}
\usepackage{extarrows}

\newtheorem{theorem}{Theorem}
\newtheorem{lemma}{Lemma}

\newtheorem{proposition}{Proposition}
\newtheorem{example}{Example}
\newtheorem{definition}{Definition}
\newtheorem{remark}{Remark}

\usepackage{xspace}
\usepackage{bbm}
\catcode`@=11
\chardef\mathlig@atcode\count255

\def\actively#1#2{\begingroup\uccode`\~=`#2\relax\uppercase{\endgroup#1~}}
\def\mathlig@gobble{\afterassignment\mathlig@next@cmd\let\mathlig@next= }
\def\mathlig@delim{\mathlig@delim}
\def\mathlig@defcs#1{\expandafter\def\csname#1\endcsname}
\def\mathlig@let@cs#1#2{\expandafter\let\expandafter#1\csname#2\endcsname}
\def\mathlig@appendcs#1#2{\expandafter\edef\csname#1\endcsname{\csname#1\endcsname#2}}

\def\mathlig#1#2{\mathlig@checklig#1\mathlig@end\mathlig@defcs{mathlig@back@#1}{#2}\ignorespaces}

\def\mathlig@checklig#1#2\mathlig@end{%
 \expandafter\ifx\csname mathlig@forw@#1\endcsname\relax
 \expandafter\mathchardef\csname mathlig@back@#1\endcsname=\mathcode`#1%
 \mathcode`#1"8000\actively\def#1{\csname mathlig@look@#1\endcsname}%
 \mathlig@dolig#1\mathlig@delim
\fi
\mathlig@checksuffix#1#2\mathlig@end
}

\def\mathlig@checksuffix#1#2\mathlig@end{%
\ifx\mathlig@delim#2\mathlig@delim\relax\else\mathlig@checksuffix@{#1}#2\mathlig@end\fi
}
\def\mathlig@checksuffix@#1#2#3\mathlig@end{%
\expandafter\ifx\csname mathlig@forw@#1#2\endcsname\relax\mathlig@dosuffix{#1}{#2}\fi
\mathlig@checksuffix{#1#2}#3\mathlig@end
}

\def\mathlig@dosuffix#1#2{%
\mathlig@appendcs{mathlig@toks@#1}{#2}%
\mathlig@dolig{#1}{#2}\mathlig@delim
}

\def\mathlig@dolig#1#2\mathlig@delim{%
 \mathlig@defcs{mathlig@look@#1#2}{%
 \mathlig@let@cs\mathlig@next{mathlig@forw@#1#2}\futurelet\mathlig@next@tok\mathlig@next}%
 \mathlig@defcs{mathlig@forw@#1#2}{%
  \mathlig@let@cs\mathlig@next{mathlig@back@#1#2}%
  \mathlig@let@cs\checker{mathlig@chck@#1#2}%
  \mathlig@let@cs\mathligtoks{mathlig@toks@#1#2}%
  \expandafter\ifx\expandafter\mathlig@delim\mathligtoks\mathlig@delim\relax\else
  \expandafter\checker\mathligtoks\mathlig@delim\fi
  \mathlig@next
 }%
 \mathlig@defcs{mathlig@toks@#1#2}{}%
 \mathlig@defcs{mathlig@chck@#1#2}##1##2\mathlig@delim{%
  \ifx\mathlig@next@tok##1%
   \mathlig@let@cs\mathlig@next@cmd{mathlig@look@#1#2##1}\let\mathlig@next\mathlig@gobble
  \fi 
  \ifx\mathlig@delim##2\mathlig@delim\relax\else
   \csname mathlig@chck@#1#2\endcsname##2\mathlig@delim
  \fi
 }%
 \ifx\mathlig@delim#2\mathlig@delim\else
  \mathlig@defcs{mathlig@back@#1#2}{\csname mathlig@back@#1\endcsname #2}%
 \fi
}%

\catcode`@\mathlig@atcode

\newcommand{\muspace}{\mspace{1mu}}

\DeclareRobustCommand{\scond}{\mathchoice{\muspace\vert\muspace}{\vert}{\vert}{\vert}}
\mathlig{|}{\scond}

\DeclareRobustCommand{\discint}{\mathchoice{\mspace{-1.5mu}:\mspace{-1.5mu}}{\mspace{-1.5mu}:\mspace{-1.5mu}}{:}{:}}
\mathlig{::}{\discint}

\newcommand{\Bc}{\mathcal{B}}

\newcommand{\Ic}{\mathcal{I}}

\newcommand{\Nc}{\mathcal{N}}

\newcommand{\Bcal}{\mathcal{B}}

\newcommand{\Gcal}{\mathcal{G}}

\newcommand{\Kcal}{\mathcal{K}}

\newcommand{\Xcal}{\mathcal{X}}

\newcommand{\Cr}{\mathscr{C}}

\newcommand{\Rr}{\mathscr{R}}

\newcommand{\Av}{{\bf A}}
\newcommand{\Cv}{{\bf C}}

\newcommand{\Xv}{{\bf X}}
\newcommand{\Pv}{{\bf P}}

\newcommand{\Rv}{{\bf R}}

\newcommand{\tv}{{\bf t}}
\newcommand{\xv}{{\bf x}}

\newcommand{\Xh}{{\hat{X}}}

\let\P\relax
\DeclareMathOperator\P{\textsf{P}}

\newcommand{\N}{\mathrm{N}}

\def\textiid{i.i.d.\@\xspace}
\newcommand\iid{\ifmmode\text{ i.i.d. } \else \textiid \fi}

\def\clap#1{\hbox to 0pt{\hss#1\hss}}
\def\mathclap{\mathpalette\mathclapinternal}
\def\mathclapinternal#1#2{%
  \clap{$\mathsurround=0pt#1{#2}$}}

\let\oldstackrel\stackrel
\renewcommand{\stackrel}[2]{\oldstackrel{\mathclap{#1}}{#2}}

\allowdisplaybreaks

\begin{document}
\title{Secure Index Coding with Security Constraints on Receivers\vspace{-2mm}}

\author{
Yucheng Liu, Parastoo Sadeghi\\      Research School of Electrical,   \\
Energy and Materials Engineering (RSEEME)\\ Australian National University, Australia\\ \{yucheng.liu,parastoo.sadeghi\}@anu.edu.au   
  \and
Neda Aboutorab, Arman Sharififar\\      School of Engineering and Information Technology\\ University of New South Wales, Australia\\ n.aboutorab@unsw.edu.au   \\ a.sharififar@student.unsw.edu.au
}

\maketitle

\begin{abstract}

Index coding is concerned with efficient broadcast of a set of messages to receivers in the presence of receiver side information. In this paper, we study the secure index coding problem with security constraints on the \emph{receivers themselves}. That is, for each receiver there is a single \emph{legitimate} message it needs to decode and a \emph{prohibited} message list, none of which should be decoded by that receiver. To this end, our contributions are threefold. We first introduce a secure linear coding scheme, which is an extended version of the fractional local partial clique covering scheme that was originally devised for non-secure index coding. We then develop two information-theoretic bounds on the performance of any valid secure index code, namely secure polymatroidal outer bound (on the capacity region) and secure maximum acyclic induced subgraph lower bound (on the broadcast rate). The structure of these bounds leads us to further develop two necessary conditions for a given index coding problem to be securely feasible (i.e., to have nonzero rates).

\end{abstract}

\section{Introduction}\label{sec:intro}


Index coding, introduced by Birk and Kol in the context of satellite communication \cite{Birk--Kol1998}, studies the efficient broadcast problem where a server broadcasts messages to multiple receivers via a noiseless channel. 
Each receiver requests one unique message and has prior knowledge of some other messages as its side information.  
Despite substantial progress achieved so far (see \cite{arbabjolfaei2018fundamentals} and the references therein), the index coding problem remains open in general. 

The index coding problem with security constraints was first studied by Dau et al. in \cite{dau2012security}, where in addition to the \emph{legitimate} receivers there is an eavesdropper who knows some messages as its side information and wants to obtain any other message. 
The server must broadcast in such a way that the legitimate receivers can decode their requested messages while the eavesdropper cannot decode any single message aside from the messages it already knows. 
Several extensions have been studied in \cite{ong2016secure,ong2018secure,mojahedian2017perfectly,liu:vellambi:kim:sadeghi:itw18}. 

In this work, we consider the index coding problem with \emph{security constraints on the legitimate receivers themselves}. That is, instead of assuming the existence of an eavesdropper, we impose security requirements on the legitimate receivers, such that each receiver must decode the legitimate message it requests and, at the same time, cannot learn any single message from a certain subset of \emph{prohibited} messages. 
Such communication model has clear applications, e.g., in video streaming scenarios. While the video streaming provider must ensure the users can get the movie they have requested and paid for, it also needs to prevent them from downloading movies that they 
have not paid for. 
Such model was introduced and briefly studied in \cite[Section IV-E]{dau2012security}, where the authors focused on the \emph{linear} index codes only. 
Moreover, a special case of this model has been studied in \cite[Section VI]{narayanan2018private}, where the prohibited message set for each receiver includes all the messages that are neither requested nor known as side information by the receiver. 
In the rest of the paper, we refer to the index coding problem with security constraints on legitimate receivers simply as secure index coding. 

The contribution of this paper are as follows.
\begin{enumerate}
\item We introduce a practical linear coding scheme (Theorem \ref{thm:s-flpcc}) for secure index coding based on the fractional local partial clique covering scheme from \cite{arbabjolfaei2014local} for non-secure index coding. In particular, to show that such coding scheme satisfies the security constraints on receivers, we make use of the existence conditions of secure index codes for the index coding problem in the presence of an eavesdropper studied in \cite{ong2016secure}. 
\item We derive two information-theoretic performance bounds (Theorems \ref{thm:outer:bound} and \ref{thm:secure:mais}) for secure index coding, which can be seen as the secure variants of the polymatroidal bound \cite{blasiak2011lexicographic, Arbabjolfaei--Bandemer--Kim--Sasoglu--Wang2013} and the maximum acyclic induced subgraph (MAIS) bound \cite{bar2011index}, respectively. 
\item We propose two necessary conditions (Theorems \ref{thm:necessary:1} and \ref{thm:necessary:2}) for a given secure index coding problem to be \emph{feasible}. This is useful, becasue for some problems there exists no index code that can simultaneously satisfy the decoding requirement and the security constraints of all receivers. We say such problems are infeasible. 

\end{enumerate}

The paper is organized as follows. In Section \ref{sec:model}, we state the problem setup and stipulate system requirements. 
In Section \ref{sec:linear}, we propose our secure linear coding scheme and its corresponding inner bound on the capacity region of the secure index coding problem. 
In Section \ref{sec:necessary}, we develop performance bounds and necessary conditions for the feasibility of the problem.
Concrete examples are provided in Sections \ref{sec:linear} and \ref{sec:necessary} to show the efficacy of the proposed techniques. 
We conclude the paper in Section \ref{sec:conclusion}.



For non-negative integers $a$ and $b$, $[a]\doteq \{1,2,\cdots,a\}$, and $[a:b]\doteq \{a,a+1,\cdots,b\}$. If $a>b$, $[a:b]=\emptyset$. For a set $S$, $|S|$ denotes its cardinality and $2^S$ denotes its power set. 

\vspace{-1mm}

\section{System Model}\label{sec:model}

\vspace{-1mm}

Assume that there are $n$ messages, $x_i \in \{0,1\}^{t_i}, i \in [n]$, where $t_i$ is the length of binary message $x_i$.  
For brevity, when we say message $i$, we mean message $x_i$. 
Let $X_i$ be the random variable corresponding to $x_i$. We assume that $X_1, \ldots, X_n$ are independent and uniformly distributed. 
For any $S\subseteq [n]$, set $S^c \doteq [n]\setminus S$, $\xv_S \doteq (x_i,i\in S)$, and $\Xv_S \doteq (X_i,i\in S)$. 
By convention, $\xv_{\emptyset} = \Xv_{\emptyset} = \emptyset$. 
Set $N\doteq 2^{[n]}$ denotes the power set of the message set $[n]$. 

A server that contains all messages $\xv_{[n]}$ is connected to all receivers via a noiseless broadcast link of normalized capacity $C = 1$. Let $y$ be the output of the server, which is a function of $\xv_{[n]}$. 
There are $n$ receivers, where receiver $i \in [n]$ wishes to obtain $x_i$ and knows $\xv_{A_i}$ as side information for some $A_i \subseteq [n]\setminus \{i\}$. The set of indices of \emph{interfering messages} at receiver $i$ is denoted by the set $B_i=(A_i\cup \{i\})^c$. 

We assume \emph{weak security} constraints against the receivers. 
That is, for each receiver $i\in [n]$, there is a set of messages $P_i\subseteq B_i$, which the receiver is prohibited from learning. 
More specifically, receiver $i$ should not be able to decode any information about each individual message $j\in P_i$ given the side information $\xv_{A_i}$ and the received codeword $y$. 
A $(\tv,r) = ((t_i, i \in [n]),r)$ {\em secure index code} is defined by
\begin{itemize}
\item An encoder at the server, $\phi: \prod_{i \in [n]} \{0,1\}^{t_i} \to \{0,1\}^{r}$, which maps the messages $\xv_{[n]}$ to an $r$-bit sequence $y$; 
\item $n$ decoders, one for each receiver $i \in [n]$, such that $\psi_i:  \{0,1\}^{r} \times \prod_{k \in A_i} \{0,1\}^{t_k} \to \{0,1\}^{t_i}$ maps the received sequence $y$ and the side information $\xv_{A_i}$ to $\hat{x}_i$.
\end{itemize}

We say a rate tuple $\mathbf{R} = (R_i, i \in [n])$ is \emph{securely} achievable if 
there exists a $(\tv,r)$ index code satisfying 
\begin{alignat}{2}
&\text{\bf Rate:}                   && \mathbin{R_i = \frac{t_i}{r}, \qquad  \forall i \in [n],}      \label{con:model:rate}    \\
&\text{\bf Message:}             && H(\Xv_S|\Xv_{S'})     \nonumber     \\
& &&=H(\Xv_S)=\sum_{i\in S}t_i,  \qquad \forall S,S'\in N,  \label{con:model:message} \\
&\text{\bf Codeword:}            && H(Y)\le r,    \label{con:model:codeword}    \\
&\text{\bf Encoding:}             && H(Y|\Xv_{[n]})=0,   \label{con:model:encoding}    \\
&\text{\bf Decoding:}   \quad  && H(X_i|Y,\Xv_{A_i})=0,    \qquad  \forall i\in [n],  \label{con:model:decoding}    \\
&\text{\bf Security:}              && I(X_j;Y|\Xv_{A_i})=0,  \qquad   \forall j\in P_i,i\in [n],     \label{con:model:security}  
\end{alignat}
where \eqref{con:model:rate} is the definition of $R_i,i\in [n]$, 
\eqref{con:model:message} follows from the assumption that the messages are independent and uniformly distributed, 
\eqref{con:model:codeword} is due to the length of the codeword being $r$, 
\eqref{con:model:encoding} follows from that $y$ is a function of $\xv_{[n]}$, 
\eqref{con:model:decoding} is stipulated by the decoding requirement at receivers: $\P\{ (\Xh_1,\ldots, \Xh_n) \ne (X_1, \ldots, X_n)\}=0$, together with Fano's Inequality, 
and \eqref{con:model:security} is stipulated by the security constraints on the receivers. 

The capacity region $\Cr$ of a given secure index coding problem is the closure of the set of all its securely achievable rate tuples $\Rv$. The symmetric capacity is defined as 
\begin{align} \label{eq:symmetric:capacity}
C_{\rm sym}=\max \{ R:(R,\cdots,R) \in \Cr \}.
\end{align}
The broadcast rate $\beta$, which characterizes the minimum number of the required transmissions from the server to satisfy the decoding and security constraints for each receiver when the messages are of the same length, is defined as 
\begin{align}
\beta=1/C_{\rm sym}. 
\end{align}

Any secure index coding problem can be represented by a tuple $(\Av,\Pv)$ where $\Av \doteq (A_i,i\in [n])$ and $\Pv \doteq (P_i,i\in [n])$ specify the side information availability and the security constraints at receivers, respectively. For example, for a three-message problem with $A_1=\emptyset, A_2=\{3\}, A_3=\{2\}$, and $P_1=\{2,3\}, P_2=P_3=\emptyset$, we write
\begin{align}
(\Av,\Pv)=( (\emptyset,\{3\},\{2\}) , (\{2,3\},\emptyset,\emptyset) ).  \label{eq:toy:example}
\end{align}
Given a secure index coding problem $(\Av,\Pv)$, its non-secure counterpart with the same $\Av$ and no security constraints can be denoted as $(\Av,(\emptyset,\cdots,\emptyset))$, or simply as $\Av$. 

Any $n$-message non-secure index coding problem can be equivalently represented by its side information graph $\Gcal$ with $n$ vertices, in which vertex $i\in [n]$ represents message $i$ and a directed edge $(i,j)$ means that $i\in A_j$. 
Therefore, any secure index coding problem $(\Av,\Pv)$ can also be represented by the tuple $(\Gcal,\Pv)$, where $\Gcal$ represents the side information graph corresponding to $\Av$. 
For a given $\Gcal$, for any set $S\in N$, $\Gcal|_S$ denotes the subproblem/subgraph of $\Gcal$ induced by $S$. 


We denote the capacity region and the broadcast rate of the problem $(\Gcal,\Pv)$ as $\Cr(\Gcal,\Pv)$ and $\beta(\Gcal,\Pv)$, respectively, when such dependence is to be emphasized. 

\vspace{-0.5mm}
\section{A Secure Linear Coding Scheme}   \label{sec:linear}
\vspace{-0.5mm}

In this section, we extend the fractional local partial clique covering (FLPCC) coding scheme from \cite{arbabjolfaei2014local} to the secure index coding problem. 
First, we briefly review the results from \cite{arbabjolfaei2014local}. 
For a given non-secure index coding problem $\Gcal$ define $\kappa(\Gcal)\doteq n-\min_{i\in [n]}|A_i|$, which denotes the number of 
parity symbols to be transmitted 
if one applies a maximum distance separable (MDS) code to the problem. Subsequently, for any $J \in N$ and its induced subproblem $\Gcal|_J$, $\kappa(\Gcal|_J)=|J|-\min_{i\in J}|A_i\cap J|$. 


The FLPCC scheme applies time sharing among a number of subproblems $\Gcal|_J$ for some $J \in N$, where each subproblem $\Gcal|_J$ is assigned with a certain fraction of the unit channel capacity $\lambda_J$. 
Any message $x_i$ that appears in multiple subproblems is split into sub-messages $x_{i,J},J\in \{ J\in N:i\in J,\lambda_J>0 \}$, and 
for each subproblem $\Gcal|_J$, a systematic $(|J|+\kappa(\Gcal|_J),|J|)$ MDS code is used such that every receiver $i\in J$ can decode sub-message $x_{i,J}$ at rate $\frac{\lambda_J}{\kappa(\Gcal|_J)}$. 
Moreover, an MDS code is applied at the subproblem level to the MDS parity symbols for subproblems, as each receiver can recover some parity symbols from its side information. In this way, the channel capcity is shared only among the parity symbols not available locally at each receiver. 

The following proposition presents the inner bound on the capcacity region given by the FLPCC scheme. 

\begin{proposition}[Arbabjolfaei and Kim \cite{arbabjolfaei2014local}]  \label{propo:flpcc}
Consider a given non-secure index coding problem $\Gcal$. Its capacity region $\Cr(\Gcal)$ is inner bounded by the rate region $\Rr_{\rm FLPCC}(\Gcal)$ that consists of all rate tuples $\Rv=(R_i,i\in [n])$ such that 
\begin{align}
&R_i\le \sum_{J\in N:i\in J}\frac{\lambda_J}{\kappa(\Gcal|_J)},   \label{eq:flp:goal}   
\end{align}
for some $\lambda_J,J\in N$ satisfying that 
\begin{alignat}{2}
%
&\lambda_J\in [0,1],          &&     \qquad   \forall J\in N,   \label{eq:flp:1}   \\
&\sum_{J\in N:J\not \subseteq A_i}\lambda_{J}\le 1,          &&   \qquad  \forall i\in [n].    \label{eq:flp:2}  
\end{alignat}
\end{proposition}


Now we describe our extended secure fractional local partial clique covering (S-FLPCC) scheme for the secure index coding problem $(\Gcal,\Pv)$. 
The S-FLPCC scheme still utilizes time sharing and rate splitting among a number of subproblems of $\Gcal$ and applies an MDS code at the subproblem level while each subproblem also uses an MDS code. 
The main difference with the FLPCC scheme is that for the S-FLPCC scheme we consider only the subproblems that satisfy relevant security constraints. 

More specifically, for each subproblem $\Gcal|_J$ assigned with a nonzero fraction of the channel capacity $0<\lambda_J\le 1$, we use a systematic $(|J|+\kappa(\Gcal|_J),|J|)$ MDS code such that every receiver $i\in J$ can decode sub-message $x_{i,J}$ of rate $\frac{\lambda_J}{\kappa(\Gcal|_J)}$. 

For each subproblem $\Gcal|_J$, due to the nature of MDS codes, every receiver $i\in J$ will be able to decode all the sub-messages in $J$ of rate $\frac{\lambda_J}{\kappa(\Gcal|_J)}$ from the corresponding parity symbols. Hence, we require that $J\cap P_i=\emptyset$, since otherwise receiver $i$ will be able to obtain some information about the messages in $J\cap P_i$, which violates the security constraint \eqref{con:model:security}. 

On the other hand, any receiver $i\notin J$ acts like an eavesdropper to the subproblem $\Gcal|_J$ if $J\cap P_i\neq \emptyset$. 
It has been shown in \cite{ong2016secure} that for $\Gcal|_J$ there exists some systematic  $(|J|+\kappa(\Gcal|_J),|J|)$ MDS code over a large enough finite field that is secure against an eavesdropper who knows up to $|J|-\kappa(\Gcal|_J)$ messages within $\Gcal|_J$ as its side information (see \cite[Theorem 1]{ong2016secure} and its proof for more details). 
Therefore, to make sure that receiver $i\notin J$ with $J\cap P_i\neq \emptyset$ cannot learn any single message from the parity symbols of $\Gcal|_J$, we simply require that $|A_i\cap J|< |J|-\kappa(\Gcal|_{J})$. 


Referring to our system model, sub-messages $x_{i,J},J\in \{ J\in N:i\in J,\lambda_J>0 \}$ are independent of each other. Hence, by combining parity symbols from different subproblems a receiver cannot gain any extra information than considering the parity symbols for each subproblem separately. Therefore, the general security constraint in \eqref{con:model:security} can be satisfied as long as the aforementioned security constraints for each subproblem are satisfied. 


%
%

We present the S-FLPCC inner bound on the capacity region, $\Cr_{\rm S-FLPCC}(\Gcal,\Pv)$, achievable by the S-FLPCC scheme. 

\begin{theorem}  \label{thm:s-flpcc}
Consider a given secure index coding problem $(\Gcal,\Pv)$. Its capacity region $\Cr(\Gcal,\Pv)$ is inner bounded by the rate region $\Rr_{\rm S-FLPCC}(\Gcal,\Pv)$ that consists of all rate tuples $\Rv=(R_i,i\in [n])$ such that 
\begin{align}
&R_i\le \sum_{J\in N:i\in J}\frac{\lambda_J}{\kappa(\Gcal|_J)},   \label{eq:s-flp:goal}   
\end{align}
for some $\lambda_J,J\in N$ satisfying
\begin{align}
&\lambda_J\in [0,1],             \qquad   \forall J\in N,   \label{eq:s-flp:1}   \\
&\sum_{J\in N:J\not \subseteq A_i}\lambda_{J}\le 1,            \qquad  \forall i\in [n],    \label{eq:s-flp:2}   \\
&P_i\cap J=\emptyset,                                       \qquad   \forall J\in N, \lambda_J>0, i\in J,  \label{eq:s-flp:3}   \\
&\text{$P_i\cap J=\emptyset$ or $|A_i\cap J|< |J|-\kappa(\Gcal|_{J})$},     \nonumber   \\
& \qquad \qquad \forall J\in N, \lambda_J>0, i\notin J.    \label{eq:s-flp:4}  
\end{align}
\end{theorem}


\begin{remark}
Note that \eqref{eq:s-flp:goal}-\eqref{eq:s-flp:2} together form the same achievable rate region for the FLPCC scheme for non-secure index coding in Proposition \ref{propo:flpcc}. The security constraints against receivers are enforced by \eqref{eq:s-flp:3} and \eqref{eq:s-flp:4}. 
\end{remark}

The example below shows the efficacy of the S-FLPCC scheme. For simplicity, we compute the securely achievable symmetric rate rather than the whole rate region. 
\begin{example}   \label{exm:s-flpcc}
Consider the following $9$-message secure index coding problem $(\Av,\Pv)$ with $P_i=B_i$ for any $i\in [9]$,
\begin{align*}
\Big(  \quad \big( &\{1\}^c, \{2\}^c, \{ 4,5,6,8,9 \}, \{ 5,6,7,8 \}, \{ 3,4,7,8,9 \},   \\
&\quad \{ 2,3,4,5,7,9 \}, \{7\}^c, \{8\}^c, \{9\}^c  \big),     \\
\big( & \emptyset, \emptyset, \{1,2,7\}, \{1,2,3,9\}, \{1,2,6\}, \{1,8\}, \emptyset, \emptyset, \emptyset \big) \quad \Big).
\end{align*}
The symmetric rate $R=\frac{1}{4}$ can be securely achieved by assigning $\lambda_{J}=\frac{1}{4}$ to the subproblems $\Gcal|_J$ for $J\in \{ \{1,2,8\}, \{2,6,7,9\}, \{3,9\}, \{4,5\} \}$, which is optimal for this problem (see Example \ref{exm:secure:mais} in Section \ref{sec:necessary} for more details). For each subproblem $\Gcal|_J$, we have $\kappa(\Gcal|_J)=1$ (i.e., the induced subgraph $\Gcal|_J$ is actually a clique). One can check that the security constraints \eqref{eq:s-flp:3} and \eqref{eq:s-flp:4} are met for each subproblem. For example, consider $J=\{2,6,7,9\}$. For any $i\in J$, $P_i\cap J=\emptyset$ and thus \eqref{eq:s-flp:3} is satisfied for this $J$. The receivers $i\in J^c=\{ 1,3,4,5,8 \}$ can be divided into two groups. For $i\in \{1,8\}$, we have $P_i=\emptyset$ and thus $P_i\cap J=\emptyset$. For $i\in \{3,4,5\}$, we have $|A_i\cap J|=2<|J|-\kappa(\Gcal|_J)=3$. Therefore, \eqref{eq:s-flp:4} is also satisfied for this $J$. 

%
\end{example}



\section{Performance Bounds and Necessary Conditions for Feasibility}  \label{sec:necessary}

In this section, we introduce two performance bounds for the secure index coding problem, as well as two necessary conditions for a given problem to be securely feasible. 

\subsection{An Outer Bound on the Capacity Region}

\begin{theorem}   \label{thm:outer:bound}
Consider a given secure index coding problem $(\Gcal,\Pv)$ with any valid $(\tv,r)$ secure index code. 
If a rate tuple $\Rv=(R_i,i\in [n])$ is securely achievable, then it must satisfy
\begin{align}
R_i=g(B\cup \{i\})-g(B),   \quad \forall B\subseteq B_i,i\in [n],\label{eq:Ri}
\end{align}
for some set function $g(S),S\in N$, such that
\begin{align}
&g(\emptyset)=0,    \\
&g([n])\le 1,     \\
&g(S)\le g(S'),    \qquad    \text{if $S\subseteq S'$},  \label{eq:monotone}  \\
&g(S\cap S')+g(S\cup S')\le g(S)+g(S'),    \label{eq:submod} \\
&g(B_i)=g(B_i\setminus \{j\}),    \qquad    \forall j\in P_i,i\in [n]. \label{eq:securitynew} 
\end{align}
\end{theorem}

\begin{IEEEproof}
Define the set function as 
\begin{align}
g(S)\doteq \frac{1}{r}H(Y|\Xv_{S^c}),  \quad \forall S\in N.    \label{eq:g:def}
\end{align}
Properties \eqref{eq:Ri}-\eqref{eq:submod} are derived in the same manner as \cite[Section 5.2]{arbabjolfaei2018fundamentals} according to the system model conditions \eqref{con:model:rate}-\eqref{con:model:decoding}.  
To show \eqref{eq:securitynew}, for any $i\in [n]$, $j\in P_i$, by \eqref{con:model:message}, \eqref{con:model:decoding}, \eqref{eq:g:def}, as well as the security constraints specified in \eqref{con:model:security}, 
we have 
\begin{align}
rg(B_i)&=H(Y|\Xv_{A_i\cup \{i\}})      \nonumber    \\
&=H(X_i|Y,\Xv_{A_i})+H(Y|\Xv_{A_i})-H(X_i|\Xv_{A_i})            \nonumber    \\
&=H(Y|\Xv_{A_i})-I(X_j;Y|\Xv_{A_i})-H(X_i|\Xv_{A_i})            \nonumber    \\
&=H(Y|\Xv_{A_i\cup \{j\}})-H(X_i|\Xv_{A_i\cup \{j\}})            \nonumber    \\
&=H(Y|\Xv_{A_i\cup \{i\} \cup \{j\}})-H(X_i|Y,\Xv_{A_i\cup \{j\}})            \nonumber    \\
&=rg(B_i\setminus \{j\}),     
\end{align}
where the second and the second last equalities are simply due to the chain rule. 
\end{IEEEproof}

The set function $g$ defined above will play a crucial role in the results to be developed henceforth.  
We particularly remark that in \eqref{eq:Ri}, when $B=\emptyset$, $g(B)=0$, and thus $R_i=g(\{ i \})$.




\subsection{A Partition of the Power Set $N$ and A Necessary Condition for Feasibility Based on It}

The security property \eqref{eq:securitynew} enforces the value of the set function $g$ to be equal for certain arguments. 
For the toy example in \eqref{eq:toy:example}, $B_1=P_1 = \{2,3\}$. Thus, by \eqref{eq:securitynew}, we have $g(\{2,3\})=g(\{2\})=g(\{3\})$. 

Moreover, combining properties \eqref{eq:monotone} and \eqref{eq:submod} of $g$ with \eqref{eq:securitynew} may result in $g$ to be equal for even more arguments. In the above example, since $g(\{2,3\})=g(\{2\})$, by \eqref{eq:submod} we have $g(\{1,2,3\})\le g(\{1,2\})+g(\{2,3\})-g(\{2\})=g(\{1,2\})$, and by \eqref{eq:monotone} we have $g(\{1,2,3\})\ge g(\{1,2\})$. Thus $g(\{1,2,3\})=g(\{1,2\})$. Similarly, $g(\{1,2,3\})=g(\{1,3\})$.

Based on the above ideas, we now formally define a partition on the set $N$, namely the $\rm g$-partition, denoted by $\Nc=\{ N_1,N_2,\cdots,N_{\gamma} \}$. 
\begin{definition}   \label{def:gpartition}
Given a secure index coding problem $(\Gcal,\Pv)$, its $g$-partition $\Nc$ can be constructed using the following steps:
\begin{enumerate}
\item \label{step:gpartition:1} Initialize the partition of $\Nc$ such that for any receiver $i\in [n]$, whose $P_i$ is nonempty, for any $T\subseteq B_i^c$, there exists a message subset $N(i,T)\in \Nc$ as 
\begin{align*}
N(i,T)=\{ T\cup B_i\setminus \{j\}:j\in P_i \} \cup \{ T\cup B_i \}.
\end{align*} 
All elements $S\in N$ that are not in any subset $N(i,T)$ are placed in $N_{\rm remaining}$, i.e.
\begin{align*}
N_{\rm remaining}=N\setminus (\cup_{T\subseteq B_i^c,i\in [n]:P_i\neq \emptyset}N(i,T)). 
\end{align*}
\item As long as there exist two subsets $N(i,T),N(i',T')$ such that $N(i,T)\cap N(i',T')\neq \emptyset$, 
we merge these two subsets into one new subset. 
We keep merging overlapping subsets until all subsets in $\Nc$ are disjoint, 
then we index the elements of $\Nc$ as $N_1,N_2,\cdots,N_{\gamma}$ in an arbitrary order, except for $N_{\gamma}=N_{\rm remaining}$. 
\label{step:gpartition:2}
\end{enumerate}
\end{definition}

For a given secure index coding problem, its $\rm g$-partition is unique. 
We call any subset within the $\rm g$-partition except for the last one a $\rm g$-subset. The function values for $g$  with arguments from within one $\rm g$-subset are always equal, enforced by \eqref{eq:monotone}, \eqref{eq:submod}, and \eqref{eq:securitynew}.

\begin{lemma}    \label{lem:gpartition}
Consider a given secure index coding problem $(\Gcal,\Pv)$ with $\rm g$-partition $\Nc=\{ N_1,N_2,\cdots,N_{\gamma} \}$ and any valid $(\tv,r)$ secure index code. For any $\rm g$-subset $N_k,k\in [\gamma-1]$, we have $$ g(S)=g(S'),    \qquad \forall S,S'\in N_k. $$
\end{lemma}
\begin{IEEEproof}
Since Step \ref{step:gpartition:2} in Definition \ref{def:gpartition} is simply merging subsets that have at least one common element, it suffices to show that for the initial partition $\Nc$ in Step \ref{step:gpartition:1}, for any receiver $i\in [n]$ whose $P_i$ is nonempty, $T\subseteq B_i^c$, we have 
\begin{align}
g(S)=g(S'), \qquad \forall S,S'\in N(i,T),    \label{eq:proof:gpartition:goal}
\end{align}


Consider any receiver $i\in [n]$ whose $P_i$ is nonempty, $T\subseteq B_i^c$. For any $j\in P_i$, we have 
\begin{align}
g(T\cup B_i)&\le g(T\cup B_i\setminus \{j\})+g(B_i)-g(B_i\setminus \{j\})   \nonumber  \\
&=g(T\cup B_i\setminus \{j\}),   \label{eq:proof:gpartition:1}
\end{align}
where the inequality follows from \eqref{eq:submod} and the equality follows from \eqref{eq:securitynew}. On the other hand, by \eqref{eq:monotone}, we have
\begin{align}
g(T\cup B_i)\ge g(T\cup B_i\setminus \{j\}).    \label{eq:proof:gpartition:2}
\end{align}
Combining \eqref{eq:proof:gpartition:1} and \eqref{eq:proof:gpartition:2} yields $g(T\cup B_i)=g(T\cup B_i\setminus \{j\})$, which implies \eqref{eq:proof:gpartition:goal}. 
\end{IEEEproof}

Let $g_{N_k}$ denote the value of $g$ of any set $S$ that belong to the $\rm g$-subset $N_k$, $k\in [\gamma-1]$, within a given $\Nc=\{ N_1,N_2,\cdots,N_{\gamma} \}$, i.e., $g_{N_k}\doteq g(S), \forall S\in N_k. $


We state our first necessary condition for a given secure index coding problem to be feasible based on its $\rm g$-partition. 

\begin{theorem}\label{thm:necessary:1}
Consider a given secure index coding problem $(\Gcal,\Pv)$ with $\rm g$-partition $\Nc=\{ N_1,N_2,\cdots,N_{\gamma} \}$. For any $k\in [\gamma-1]$, if there exist some $S,S'\in N_k$ and $i\in [n]$ such that $S'\cup \{i\}\subseteq S$ and $S'\subseteq B_i$, then the problem is infeasible. 
\end{theorem}

We show the above theorem using Lemma \ref{lem:gpartition} below. 

\begin{IEEEproof}
Consider any valid $(\tv,r)$ secure index code and any securely achievable $\Rv$. 
For any $k\in [\gamma-1]$, suppose that there exist such $S,S'\in N_k$ and $i\in [n]$. We have
\begin{align}
g(S')&=g(S)\ge g(S',i)=g(S')+R_i,   \label{eq:proof:necessary:1}
\end{align}
where the first equality follows from Lemma \ref{lem:gpartition} with that $S$ and $S'$ belong to the same $\rm g$-subset $N_k$, 
the inequality follows from that $S'\cup \{i\} \subseteq S$ and \eqref{eq:monotone}, 
and the second equality follows from \eqref{eq:Ri} with $S'\subseteq B_i$. 
Clearly, \eqref{eq:proof:necessary:1} implies that $R_i=0$ and thus the problem is infeasible. 
\end{IEEEproof}

\begin{remark}
One common scenario where a problem is infeasible is that there exit two receivers $i\neq j\in [n]$ such that $A_j\subseteq A_i\cup \{i\}$ and $j\in P_i$. For example, see \cite[Proposition 2]{narayanan2018private}. In this case, receiver $i$, after decoding its requested message $i$, knows more messages than receiver $j$ and thus can always mimic the behaviour of receiver $j$ to decode message $j$. This violates the security constraint $j\in P_i$ and thus, the problem is infeasible. 
We can simply show that such scenario is captured by Theorem \ref{thm:necessary:1} as a special case. 
Since $j\in P_i\subseteq B_i$, there exists a $\rm g$-subset $N_k$ for some $k\in [\gamma-1]$ within $\Nc$ such that $S=B_i\in N_k$ and $S'=B_i\setminus \{j\} \in N_k$. 
First, we have $$S'\cup \{j\}=(B_i\setminus \{j\}) \cup \{j\}=B_i\subseteq B_i=S. $$
Second, since $A_j\subseteq A_i\cup \{I\}$, we have $B_j\cup \{j\}=A_j^c\supseteq (A_i\cup \{i\})^c=B_i$, which together with $j\in B_i$ leads to $$S'=B_i\setminus \{j\}\subseteq B_j. $$ 
Therefore, according to Theorem \ref{thm:necessary:1}, the problem is infeasible. 
\end{remark}

\begin{example}   \label{exm:infeasible:by:necessary:condition:1}
Consider the following $5$-message secure index coding problem $(\Av,\Pv)$,
\begin{align*}
\big( &( \{2,4,5\},\{1,5\},\emptyset,\{2\},\{1,2\} )  , ( \emptyset, \{4\}, \{1,2,5\}, \{1\}, \emptyset ) \big).
\end{align*}
By Definition \ref{def:gpartition}, the $\rm g$-partition of the problem can be written as $\Nc=\{ N_1,\cdots,N_{\gamma} \}$ where $\gamma=6$, and
\begin{align*}
N_1&=\{ \{3\},\{3,4\} \}, \enskip N_2=\{ \{1,2,4,5\} \setminus \{j\}:j\in \{1,2,5\} \},  \\
N_3&=\{ \{1,3\},\{1,3,4\} \}, \enskip N_4=\{ \{2,3\},\{2,3,4\} \}  \\
N_5&=\{ \{1,2,3\}, \{1,2,3,4\}, \{3,5\}, \{1,3,5\}, \{2,3,5\},  \\
& \{1,2,3,5\}, \{3,4,5\}, \{1,3,4,5\}, \{2,3,4,5\}, [5]  \},
\end{align*}
and $N_6=N\setminus (\cup_{k\in [5]}N_{k})$. Consider $\{1,3,5\},\{1,3,4,5\}\in N_5$ and $4\in [5]$, we have $\{1,3,5\} \cup \{4\} \subseteq \{1,3,4,5\}$ and $\{1,3,5\} \subseteq B_4$. Thus, by Theorem \ref{thm:necessary:1} the problem is infeasible. 
\end{example}



\subsection{A Lower Bound on the Broadcast Rate and A Necessary Condition for Feasibility Based on It}

%
%

First, we briefly review the MAIS bound from \cite{bar2011index}. 
\begin{proposition}[Bar-Yossef et al. \cite{bar2011index}]         \label{prop:mais}
Consider a non-secure index coding problem $\Gcal$. Its broadcast rate $\beta(\Gcal)$ is lower bounded as 
\begin{align*}
\beta(\Gcal)\ge \beta_{\rm MAIS}(\Gcal) \doteq \max_{S\in N:\text{$\Gcal|_{S}$ is acyclic}}|S|. 
\end{align*}
\end{proposition}

We have the following lemma.
\begin{lemma}    \label{lem:g:mais}
Consider a non-secure index coding problem $\Gcal$ with any valid $(\tv,r)$ secure index code. For any set $S\in N$ and securely achievable symmetric rate $R$, we have
\begin{align*}
g(S)\ge R\cdot \beta_{\rm MAIS}(\Gcal|_S).
\end{align*}
\end{lemma}
\begin{IEEEproof}
Assume $u=\beta_{\rm MAIS}(\Gcal|_S)$. Then there exists some set $U=\{i_1,i_2,\cdots,i_u\} \subseteq S$ whose induced subgraph $\Gcal|_U$ is acyclic. Hence, without loss of generality, we have
\begin{align}
\{ i_1,i_2,\cdots,i_{p-1} \} \subseteq B_{i_p},  \qquad \forall p\in [u].    \label{eq:lem:g:mais:acyclic}
\end{align}
Therefore, we have
\begin{align}
g(S)&\ge g(\{ i_1,i_2,\cdots,i_u \})    \nonumber    \\
&=g(i_1,i_2,\cdots,i_{u-1})+R_{i_u}    \nonumber   \\
&=g(i_1,i_2,\cdots,i_{u-2})+R_{i_{u-1}}+R_{i_u}    \nonumber   \\
&\enskip \vdots    \nonumber   \\
&=\sum_{p\in [u]}R_{i_p}=R\cdot \beta_{\rm MAIS}(\Gcal|_S),   \nonumber
\end{align}
where the inequality follows from \eqref{eq:monotone}, 
the equalities except for the last one follow from \eqref{eq:Ri} with \eqref{eq:lem:g:mais:acyclic}, 
and the last equality simply follows from the definition of $u$. 
\end{IEEEproof}

Proposition \ref{prop:mais} and Lemma \ref{lem:g:mais} can be trivially extended to the secure index coding problem, based upon which we propose a new performance bound, namely, the secure maximum acyclic induced subgraph (S-MAIS) lower bound as follows. 


\begin{theorem}   \label{thm:secure:mais}
Consider a given secure index coding problem $(\Gcal,\Pv)$ with $\rm g$-partition $\Nc=\{ N_1,N_2,\cdots,N_{\gamma} \}$. The S-MAIS lower bound $\beta_{\rm S-MAIS}(\Gcal,\Pv)$ on its broadcast rate can be constructed by the following steps:
\begin{enumerate}
\item \label{step:secure:mais:1} For any subset $N_k$, $k\in [\gamma]$, initialize $\rho^{N_k}$ as 
$$ \rho^{N_k}=\max_{S\in N_k}\beta_{\rm MAIS}(\Gcal|_{S}). $$   
\item As long as there exist two $\rm g$-subsets $N_k,N_{\ell}$, $k\neq \ell \in [\gamma-1]$ such that there exist some sets $S\in N_k$, $S'\in N_{\ell}$ satisfying that $S'\subseteq S$, 
and that
$$\beta_{\rm MAIS}(\Gcal|_{\{ j\in S\setminus S':S' \subseteq B_j \}})+\rho^{N_{\ell}}>\rho^{N_{k}},$$ 
update $\rho^{N_{k}} \leftarrow \beta_{\rm MAIS}(\Gcal|_{\{ j\in S\setminus S':S' \subseteq B_j \}})+\rho^{N_{\ell}}$.    \label{step:secure:mais:2}
\item If no such $N_k,N_{\ell}$ exist, set $\beta_{\rm S-MAIS}(\Gcal,\Pv)=\max_{k\in [\gamma]}\rho^{N_k}$ and terminate the algorithm.     \label{step:secure:mais:3}
\end{enumerate}
\end{theorem}


\begin{IEEEproof}
We show that $\beta_{\rm S-MAIS}(\Gcal,\Pv)\le \beta(\Gcal,\Pv)$, which is equivalent to showing that $1\ge R\cdot \beta_{\rm S-MAIS}(\Gcal,\Pv)=\max_{k\in [\gamma]}\rho^{N_k}$ for any valid $(\tv,r)$ secure index code and any securely achievable symmetric rate $R$. 

If $\max_{k\in [\gamma]}\rho^{N_k}=\rho^{N_{\gamma}}$, 
as $\rho^{\N_{\gamma}}$ remains unchanged since its initialization, 
we have 
\begin{align*}
R\cdot \beta_{\rm S-MAIS}(\Gcal,\Pv)=R\cdot \rho^{N_{\gamma}}&=R\cdot \max_{S\in N_{\gamma}}\beta_{\rm MAIS}(\Gcal|_S)       \\
&\le R\cdot \beta_{\rm MAIS}(\Gcal)\le 1. 
\end{align*}

It remains to show that $1\ge R\cdot \beta_{\rm S-MAIS}(\Gcal,\Pv)$ when $\beta_{\rm S-MAIS}(\Gcal,\Pv)=\max_{k\in [\gamma]}\rho^{N_k}=\rho^{N_k}$ for some $k\in [\gamma-1]$. 
Recall that $g(S)\le 1,\forall S\in N$. 
We show that $1\ge R\cdot \beta_{\rm S-MAIS}(\Gcal,\Pv)$ by showing a slightly stronger statement that 
\begin{align}
g_{N_k}\ge R\cdot \rho^{N_k},    \qquad    \forall S\in N_k,k\in [\gamma-1].    \label{eq:proof:secure:mais:goal}
\end{align}

By induction, it suffices to show that 
\begin{enumerate}
\item for the initialized $\rho^{N_k}=\max_{S\in N_k}\beta_{\rm MAIS}(\Gcal|_{S}), k\in [\gamma-1]$, \eqref{eq:proof:secure:mais:goal} holds; 
\item for any $N_k,N_{\ell}$, $k\neq \ell \in [\gamma-1]$ satisfying the conditions in Step \ref{step:secure:mais:2} in Theorem \ref{thm:secure:mais}, the updated $\rho^{N_k}$, which equals to $\beta_{\rm MAIS}(\Gcal|_{\{ j\in S\setminus S':S' \subseteq B_j \}})+\rho^{N_{\ell}}$ for some $S\in N_k,S'\in N_{\ell}$, still satisfies \eqref{eq:proof:secure:mais:goal}, provided that $g_{N_{\ell}}\ge R\cdot \rho^{N_{\ell}}$. 
\end{enumerate}

Consider the initialized $\rho^{N_k}=\max_{S\in N_k}\beta_{\rm MAIS}(\Gcal|_{S}), k\in [\gamma-1]$. By Lemmas \ref{lem:gpartition} and \ref{lem:g:mais}, we have
\begin{align*}
g_{N_k}\ge R\cdot \beta_{\rm MAIS}(\Gcal|_S),   \qquad \forall S\in N_k,
\end{align*}
which directly leads to \eqref{eq:proof:secure:mais:goal}. 

Consider any $N_k,N_{\ell},k\neq \ell \in [\gamma-1]$ and $S\in N_k,S'\in N_{\ell}$ satisfying the conditions in Step \ref{step:secure:mais:2} in Theorem \ref{thm:secure:mais}. 
The updated $\rho^{N_{k}}=\beta_{\rm MAIS}(\Gcal|_{\{ j\in S\setminus S':S' \subseteq B_j \}})+\rho^{N_{\ell}}$.   
Set 
\begin{align}
s=\beta_{\rm MAIS}(\Gcal|_{\{ j\in S\setminus S':S' \subseteq B_j \}}).     \label{eq:proof:secure:mais:s}
\end{align}
Then, there exists some set $\{ j_1,j_2,\cdots,j_s \} \subseteq \{ j\in S\setminus S':S'\subseteq B_j \}$ whose induced subgraph is acyclic satisfying that 
\begin{align}
\{ j_1,\cdots,j_{p-1} \}\subseteq B_{j_p}, \qquad \forall p\in [s].    \label{eq:proof:secure:mais:1}
\end{align}
Note that we also have
\begin{align}
S'\subseteq B_{j_p}, \qquad \forall p\in [s],    \label{eq:proof:secure:mais:2}
\end{align}
since any $j_p$ is an element of the set $\{ j\in S\setminus S':S'\subseteq B_j \}$. Combining \eqref{eq:proof:secure:mais:1} and \eqref{eq:proof:secure:mais:2} we have
\begin{align}
S'\cup \{ j_1,\cdots,j_{p-1} \}\subseteq B_{j_p}, \qquad \forall p\in [s].    \label{eq:proof:secure:mais:3}
\end{align}
Hence, we have 
\begin{align}
g_{N_k}
&\ge g(S'\cup \{ j\in S\setminus S':S'\in B_j \})     \nonumber   \\
&\ge g(S'\cup \{j_1,j_2,\cdots,j_s\} )        \nonumber    \\
&=g(S'\cup \{j_1,j_2,\cdots,j_{s-1}\})+R_{j_s}    \nonumber    \\
&=g(S'\cup \{j_1,j_2,\cdots,j_{s-2}\})+R_{j_{s-1}}+R_{j_s}   \nonumber   \\
&\enskip \vdots   \nonumber   \\
&=g(S')+\sum_{p\in [s]}R_{j_p}     \nonumber   \\
&=g_{N_{\ell}}+R\cdot \beta_{\rm MAIS}(\Gcal|_{\{ j\in S\setminus S':S' \subseteq B_j \}})    \nonumber   \\
&\ge R\cdot (\rho^{N_{\ell}}+\beta_{\rm MAIS}(\Gcal|_{\{ j\in S\setminus S':S' \subseteq B_j \}}))=R\cdot \rho^{N_k},   \nonumber 
\end{align}
where the first and second inequalities follow from \eqref{eq:monotone}, 
the equalities except for the last two follow from \eqref{eq:Ri} with \eqref{eq:proof:secure:mais:3}, 
the second last equality follows from \eqref{eq:proof:secure:mais:s}, 
the last inequality follows from the assumption that $g_{N_{\ell}}\ge R\cdot \rho^{N_{\ell}}$, 
and, finally, the last equality follows from that the updated $\rho^{N_k}=\rho^{N_{\ell}}+\beta_{\rm MAIS}(\Gcal|_{\{ j\in S\setminus S':S' \subseteq B_j \}})$. 
\end{IEEEproof}

Now we state our second necessary condition for feasibility.


\begin{theorem}\label{thm:necessary:2}
Consider a given secure index coding problem $(\Gcal,\Pv)$ with $\rm g$-partition $\Nc=\{N_1,N_2,\cdots,N_{\gamma}\}$.
The problem is infeasible if there exists some $k\in [\gamma-1]$ such that
\begin{align*}
\rho^{N_k} > \min_{S\in N_k}|S|,
\end{align*}
where $\rho^{N_k}, k\in [\gamma-1]$ are iteratively defined by Steps \ref{step:secure:mais:1}-\ref{step:secure:mais:3} in Theorem \ref{thm:secure:mais}. 
\end{theorem}


\begin{IEEEproof}
Suppose for some $k\in [\gamma-1]$, we have $\rho^{N_k} > \min_{S\in N_k}|S|$. We show that the problem is infeasible by contradiction. 
Assume that the problem is feasible with some securely achievable symmetric rate $R>0$. There exists some $S_0\in N_k$ such that $|S_0|=\min_{S\in N_k}|S|$. Then, we have
\begin{align}
g(S_0)
&\le \sum_{i\in S_0}g(\{i\})=\sum_{i\in S_0}R_i=R\cdot \min_{s\in N_k}|S|,    \label{eq:proof:necessary:2:1}
\end{align}
where the inequality follows from repeated application of \eqref{eq:submod}, and the second last inequality follows from \eqref{eq:Ri}. 
Hence, 
\begin{align}
R\cdot \min_{S\in N_k}|S|&\ge g(S_0)     \nonumber    \\
&=g_{N_k}\ge R\cdot \rho^{N_k}>R\cdot \min_{S\in N_k}|S|,    \label{eq:proof:necessary:2:2}
\end{align}
where the first, second, and the last inequality follow from \eqref{eq:proof:necessary:2:1}, \eqref{eq:proof:secure:mais:goal}, and the assumption that $\rho^{N_k} > \min_{S\in N_k}|S|$, respectively. 
Clearly, \eqref{eq:proof:necessary:2:2} leads to a contradiction, and therefore the problem must be infeasible. 
\end{IEEEproof}

The following examples demonstrate the efficacy of Theorems \ref{thm:secure:mais} and \ref{thm:necessary:2}. 
\begin{example}   \label{exm:secure:mais}
Revisit the $9$-message secure index coding problem in Example \ref{exm:s-flpcc}. 
While the normal MAIS lower bound gives $\beta(\Gcal,\Pv) \ge \beta_{\rm MAIS}(\Gcal,\Pv)=\beta_{\rm MAIS}(\Gcal)=3$, 
the S-MAIS lower bound in Theorem \ref{thm:secure:mais} gives a strictly tighter result as 
$$ \beta(\Gcal,\Pv) \ge \beta_{\rm S-MAIS}(\Gcal,\Pv)=4>3, $$
which matches the result in Example \ref{exm:s-flpcc} and thus establishes the symmetric capacity to be $\frac{1}{4}$. 
\end{example}


\begin{example}   \label{exm:infeasible:by:necessary:condition:2}
Revisit the $5$-message secure index coding problem in Example \ref{exm:infeasible:by:necessary:condition:1}, which is infeasible according to Theorem \ref{thm:necessary:1}. 
One can also see that the problem is infeasible by Theorem \ref{thm:necessary:2} since for $N_5\in \Nc$, $\rho^{N_5}=4$, and thus $$\min_{S\in N_5}|S|\le |\{3,5\}|=2<4=\rho^{N_5}. $$
\end{example}

\section{Concluding Remarks}   \label{sec:conclusion}

In this work, we studied the secure index coding problem with security constraints against receivers. We proposed a linear coding scheme and two information-theoretic performance bounds. 
We also developed two necessary conditions for the existence of valid secure index codes that satisfy both decoding and security requirements of all receivers. 

A natural way to design more efficient coding schemes for secure index coding is to adopt more powerful schemes from the non-secure scenario. In particular, it has been shown that the recursive codes \cite{arbabjolfaei2014local} strictly outperforms the FLPCC scheme and the non-linear enhanced composite coding scheme \cite{isit:2017},\cite[Appendix 6.C]{arbabjolfaei2018fundamentals} strictly outperforms the recursive codes. How to guarantee security for these coding schemes remains to be investigated. As for the performance bounds, a series of new bounds have been recently proposed in \cite{isit:2019,allerton:2019}, which are strictly tighter than the MAIS bound and not as computationally intensive as the more general polymatroidal bound. Extending such bounds to secure index coding could be an interesting direction for future study, which may also lead to new feasibility check techniques.

\bibliographystyle{IEEEtran}
\bibliography{references}

\end{document}